\documentclass[twocolumn,preprintnumbers,amsmath,amssymb]{revtex4}
\usepackage{graphicx}
\usepackage{dcolumn}
\usepackage{bm}

\begin{document}
\title{
Terahertz  lasers  based on  optically pumped multiple
graphene structures
with 
 slot-line and dielectric waveguides 
}
\author{V.~Ryzhii\footnote{Electronic mail: v-ryzhii(at)u-aizu.ac.jp} 
}
\address{
Computational Nanoelectronics Laboratory,~University of Aizu, 
Aizu-Wakamatsu  965-8580,  and 
Japan Science and Technology Agency, CREST, Tokyo 107-0075, Japan
}
\author{ 
A.~A.~Dubinov
}
\address{
Computational Nanoelectronics Laboratory, University of Aizu, 
Aizu-Wakamatsu  965-8580,~Japan  and
Institute for Physics of Microstructures, Russian Academy of Sciences,
Nizhny Novgorod 603950, Russia 
}
\author{ 
T.~Otsuji
}
\address{
Research Institute for Electrical Communication, Tohoku University, Sendai 980-8577
and 
Japan Science and Technology Agency, CREST, Tokyo 107-0075, Japan
}
\author{ 
V.~Mitin
}
\address{Department of Electrical Engineering,
University at Buffalo, State University of New York, NY 14260, USA 
}
\author{ 
M.~S.~Shur
}
\address{Department of Electrical, Electronics, and Systems Engineering,
Rensselaer Polytechnic Institute,
Troy, NY 12180, USA 
}

\begin{abstract}
Terahertz (THz) lasers  on  
  optically pumped  multiple-graphene-layer (MGL)
 structures as their active region
are proposed  and evaluated. 
The developed device model accounts for the interband and intraband
transitions in the degenerate electron-hole plasma generated by optical
radiation
in the MGL structure
and the losses in the  slot or dielectric waveguide.
The  THz laser gain and the conditions of THz lasing
are found. It is shown that the lasers under consideration
can operate at  frequencies $\gtrsim 1$~THz  at room temperatures.
\end{abstract}

\maketitle
\newpage

\section{Introduction}
Graphene,
graphene nanoribbons, and graphene bilayers 
(see, for instance, Ref.~\cite{1})   can be used in different
terahertz (THz)  devices~\cite{2,3,4,5,6,7,8,9}.
Optical excitation of graphene can result
in the interband population inversion
and negative real part of the dynamic conductivity
of a graphene layer $\sigma_{\omega}$  in the THz range
of frequencies 
~\cite{7,8,9}.
The  negativity of the real part of the dynamic  conductivity
implies that 
the interband emission of photons
with the energy $\hbar\omega$, where $\hbar$ is the reduced Planck constant,
prevail over the inraband (Drude) absorption.   
If the THz photon losses  in  the resonant cavity 
 are
sufficiently small, the THz lasing can be realized in graphene-based
devices with optical pumping~\cite{10,11}. 
In Refs.~\cite{10,11}, the optically pumped THz lasers with a Fabri-Perot
resonator were considered.
In this paper, we propose and evaluate the THz lasers
based on multiple-graphene-layer (MGL) structures 
with  a metal slot-line    waveguide (SLW)
or a dielectric waveguide (DW)
pumped by optical
radiation.
The specific features of characteristics
of the MGL-based lasers  under consideration 
are associated with the frequency dependences of the absorption 
in the waveguides
and the gain-overlap factor, which is sensitive to the spatial distribution
of the THz electric field.
As show in the following, the  characteristics of the lasers with SLW and DW
are fairly similar, so that we will mainly focus on the device with SLW.
This in part  because the  structures with SLW exhibit somewhat better
confinement and can be used not only
for MGL-based lasers with optical pumping but also in
MGL-based injection lasers with lateral 
p-i-n junctions (both electrically induced
~\cite{12,13,14} and  formed by pertinent doping of MGL structure).

\section{Device model}

\begin{figure}[t]
\vspace*{-0.4cm}
\begin{center}
\includegraphics[width=7.0cm]{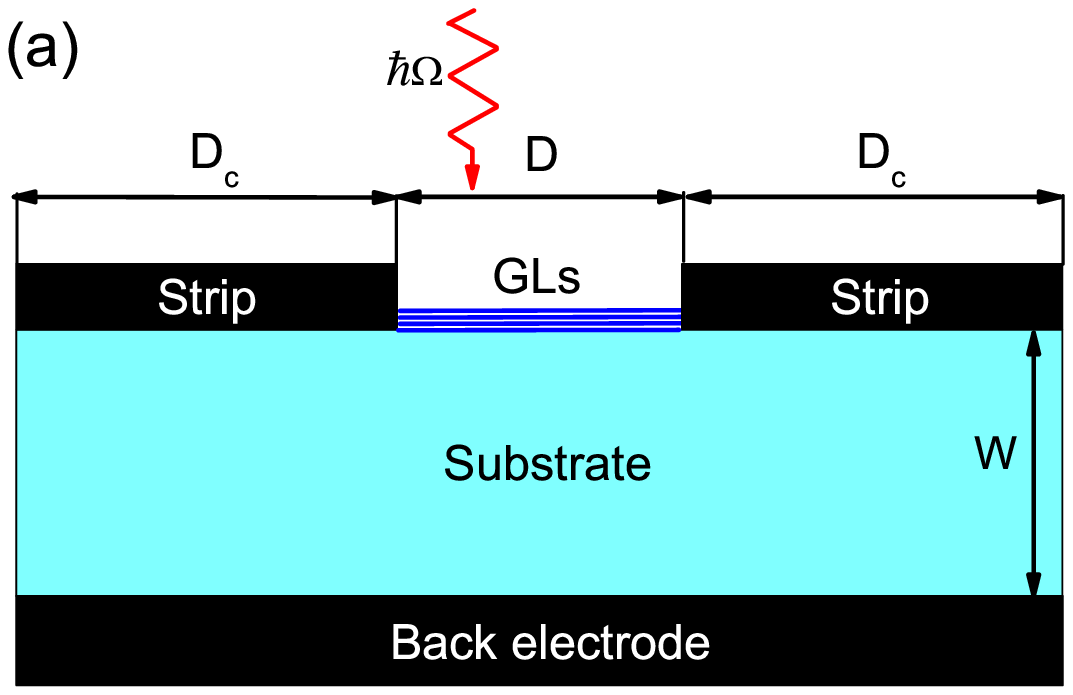}
\includegraphics[width=7.0cm]{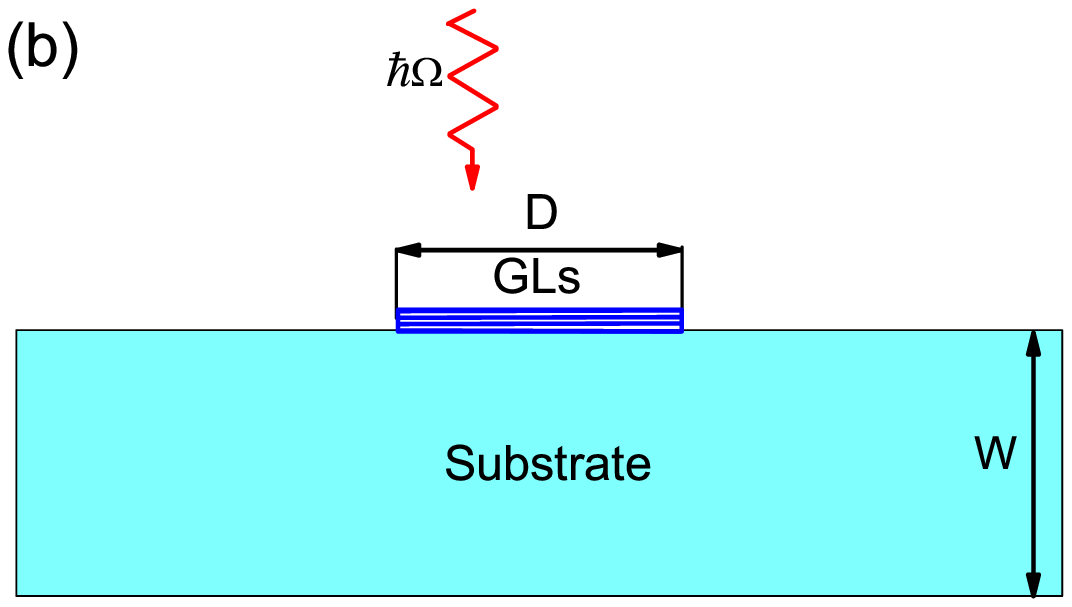}
\caption{Schematic views of the device structures under consideration:
(a) with a SLW and (b) with a DW.
}
\end{center}
\end{figure}

We consider  lasers with an MGL structure 
on a SiC or Si substrate with the side highly conducting metal
strips and a highly conducting back electrode at the substrate bottom
or on the top of  a DW. 
The cross-sections (corresponding to the $y - z$ plane),  of the 
device structures under consideration 
are  schematically shown in Fig.~1.
The axis $x$ correponds to the direction  
of the electromagnetic wave propagation,
whereas the
$y$ and $z$  directions are in the MGL structure plane
and perpendicular to it,  
respectively (see Fig.~1). The MGL  plane corresponds to $z = 0$.
The finiteness of the MGL structure thickness can be disregarded.
It is assumed that the MGL structure under consideration comprises 
$K$ upper GLs  and 
a highly 
conducting bottom GL on a SiC substrate or $K$ GLs (without the bottom GL)
  on a Si substrate.
Epitaxial MGL structures with up to $K = 100$ GLs
with very long momentum relaxation time of electrons and holes ($\tau
\simeq 20$~ps) 
were recently fabricated using the thermal
decomposition from 4H-SiC substrate~\cite{15}.
 MGL structures without the bottom GL can be fabricated using
chemical/mechanical reactions and transferred substrate techniques
which include 
chemically etching the substrate and the highly conducting bottom 
GL~\cite{16} (or mechanically 
peeling the  upper GLs) and  transferring the upper portion of
the  MGL structure on a Si or equivalent transparent substrate.
Since the  electron density and Fermi energy  $\varepsilon_F^B$ 
in the bottom GL  
is rather large
($\varepsilon_F^B   \simeq 400$~meV ~\cite{17}),
the Drude absorption in this GL can be significant
although it can be overcome by a strong emission from the upper GLs if
their number is sufficiently large. The MGL structures
without the bottom GL can exhibit significant advantages 
(see Ref.~\cite{11} and below).

It is assumed that the MGL structure is illuminated from the top
by light with the energy of photons $\hbar\Omega$. The optical waveguide
input of the pumping radiation is also possible.
When  $\hbar\Omega$ is close to
$N\hbar\omega_0/2$, where $\hbar\omega_0 \simeq 0.2$~eV 
is the optical phonon energy and $N$ is an integer,
the photogeneration of electrons and holes and 
 their cooling associated with
 the cascade emission of optical phonons,
result in an essentially occupation (population inversion) of
low energy
states near the bottom of the conduction band and 
the top of the valence band.
At  elevated
 electron and hole densities (i.e., at
 sufficiently strong optical pumping), 
the electron and hole distributions  in  the range of energies
$\varepsilon \ll \hbar\omega_0$ in the $k-$th GL 
($1 \leq k \leq K$) can be  described by
the Fermi  functions 
with the  
 quasi-Fermi energies
  $\varepsilon_F^{(k)}$.

The quasi-Fermi energies in the GLs with $k \geq 1$ are mainly determined
by the electron (hole) density in  this layer $\Sigma_k$, i.e,
 $\varepsilon_F^{(k)}  \propto \sqrt{\Sigma^{(k)}}$ and, therefore,
by the rate of photogeneration  $G_{\Omega}^{(k)}$ 
by the optical
radiation  at the $k-$th GL plane.
Considering the attenuation of the optical 
pumping radiation due to its absorption
in each GL, one can obtain
\begin{figure}[t]
\vspace*{-0.4cm}
\begin{center}
\includegraphics[width=7.0cm]{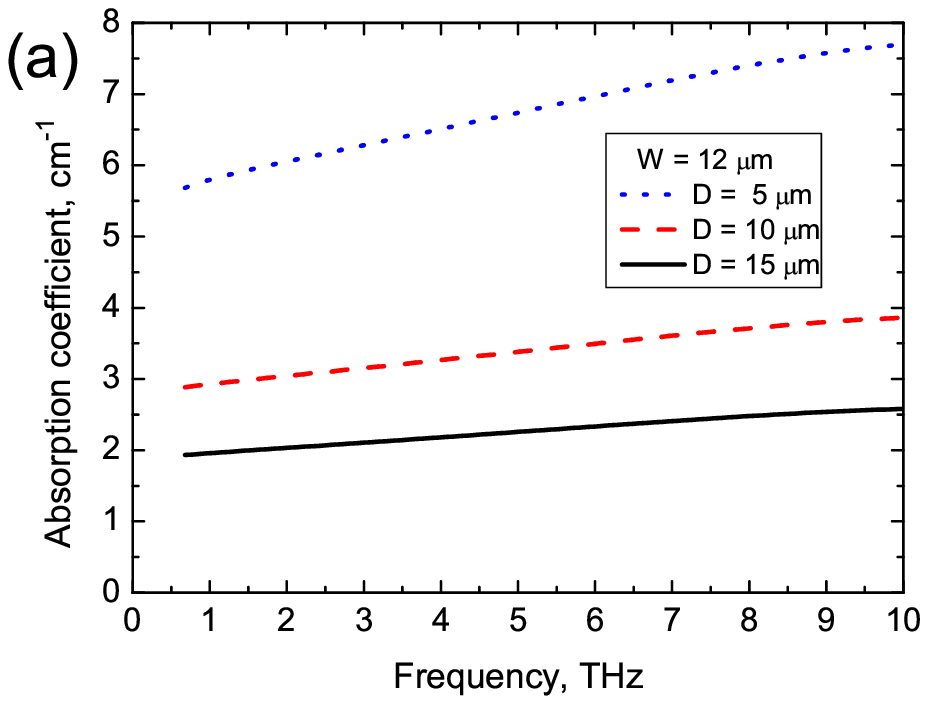}
\includegraphics[width=7.0cm]{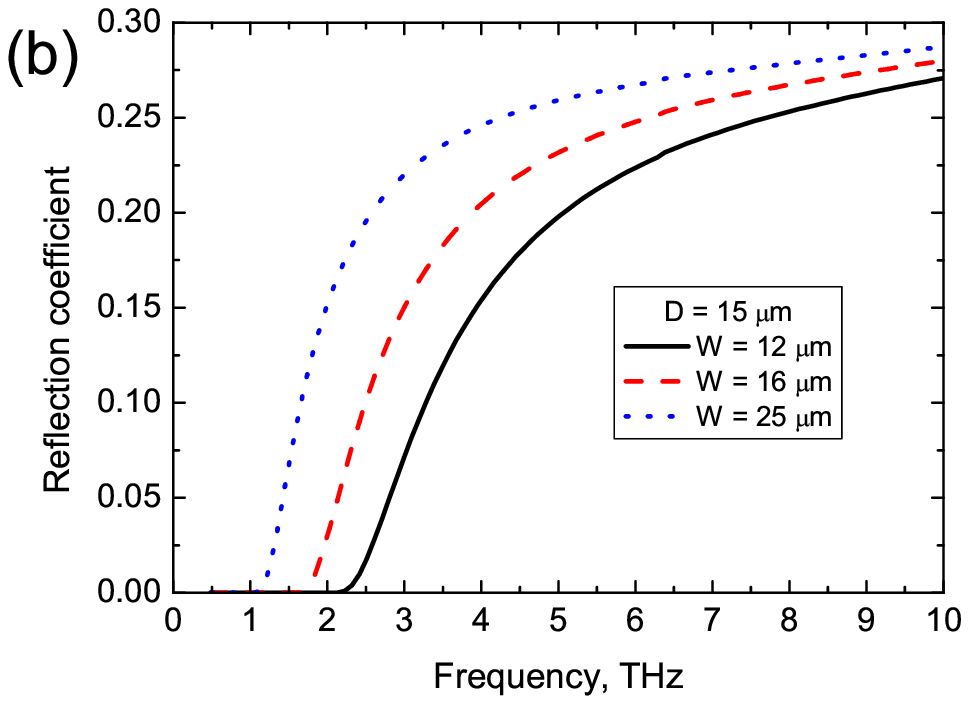}
\caption{Coefficients of (a) absorption and (b) reflection
versus frequency for SLWs with different geometrical parameters.
}
\end{center}
\end{figure}

\begin{figure}[t]
\vspace*{-0.4cm}
\begin{center}
\includegraphics[width=8.0cm]{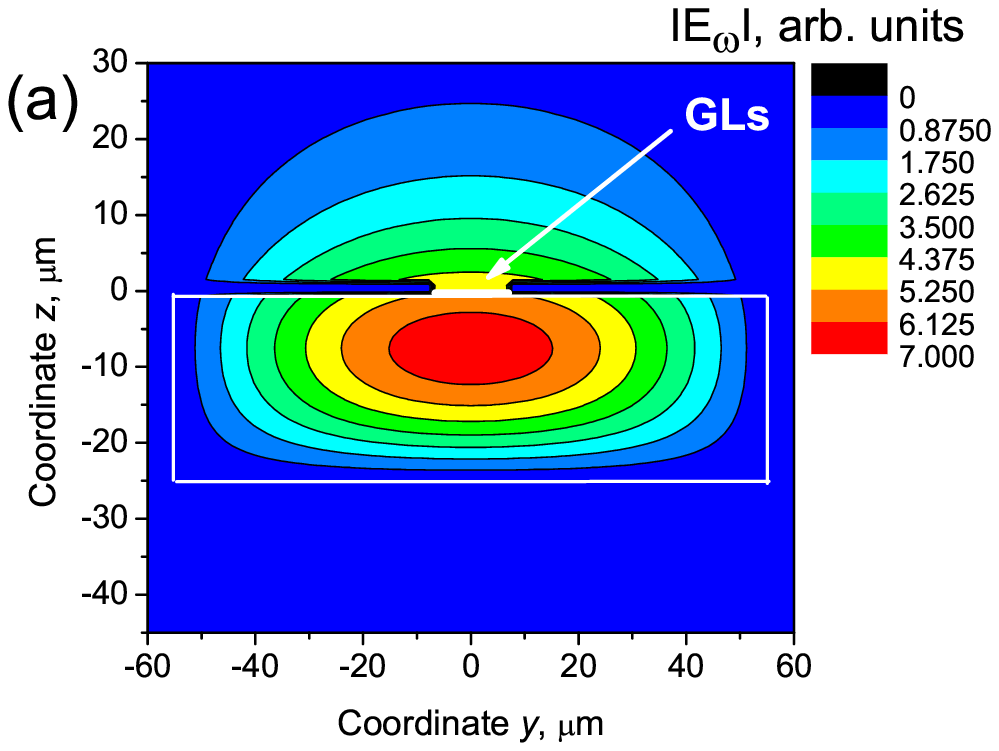}
\includegraphics[width=7.5cm]{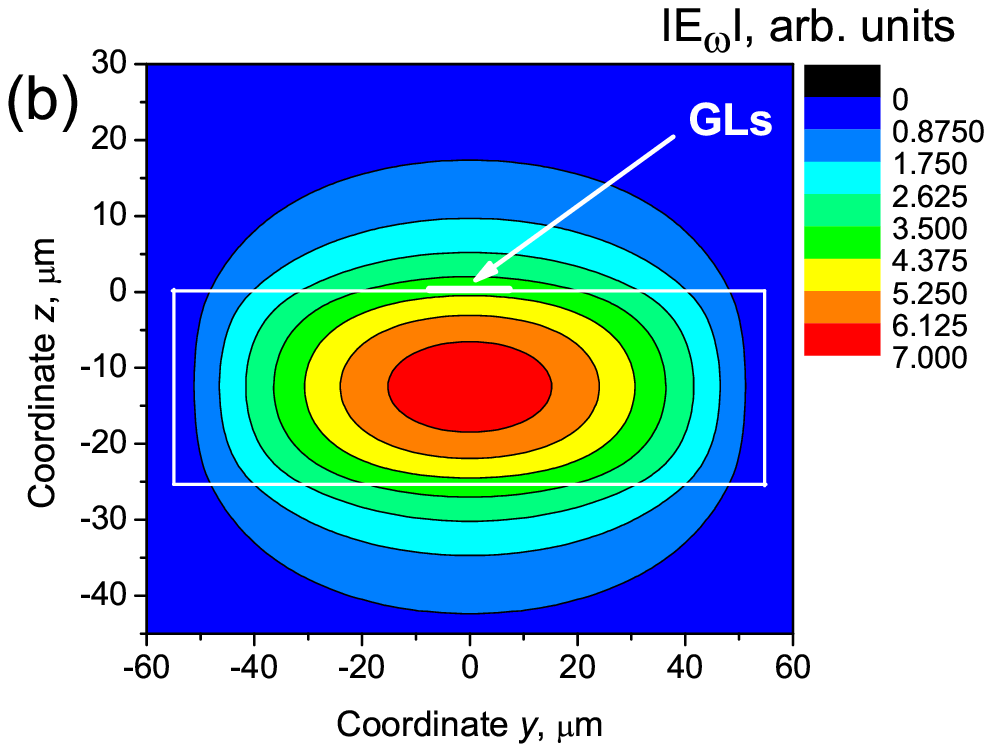}
\caption{Spatial distributions of THz electric field  (a) in SLW
cross-section
and (b) in  DW cross-section.
}
\end{center}
\end{figure}   

\begin{equation}\label{eq1}
G_{\Omega}^{(k)} =  
\frac{I_{\Omega}}{\hbar\Omega} \beta[(1 - \beta)^{K - k} + (1 - \beta_B)^2
(1 - \beta)^{K + k- 1} ].
\end{equation}
%
Here $I_{\Omega}$ is the intensity of incident pumping radiation,
$\beta = \pi\,e^2/\hbar\,c \simeq 0.023$, where $e$ is the electron charge,
$c$ is the speed of light in vacuum,
and $\beta_B = (4\pi/c) {\rm }{\rm Re}\sigma_{\Omega}^B$.
The latter quantity accounts for the absorption of optical pumping radiation
in the bottom layer. 
A relationship between $\varepsilon_F^{(k)}$ and $G_{\Omega}^{(k)}$
is determined by the recombination mechanisms~\cite{11}.
As a result,  $\varepsilon_F^{(k)}$ can be expressed via
the quasi-Fermi energy
in the topmost GL
$\varepsilon_F^{T} = \varepsilon_F^{(K)}$, which, in turn, is a function of
the  the intensity of incident pumping radiation $I_{\Omega}$.

\begin{figure}[t]
\vspace*{-0.4cm}
\begin{center}
\includegraphics[width=7.0cm]{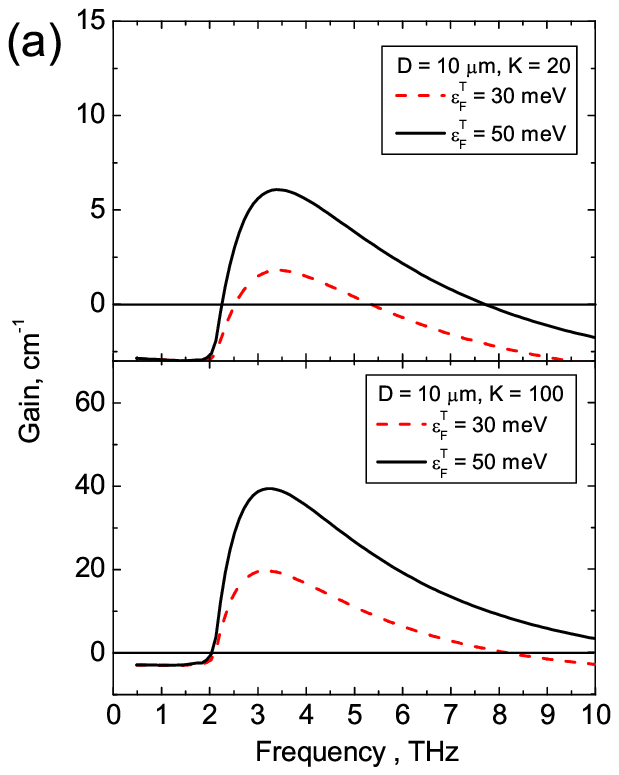}
\includegraphics[width=7.0cm]{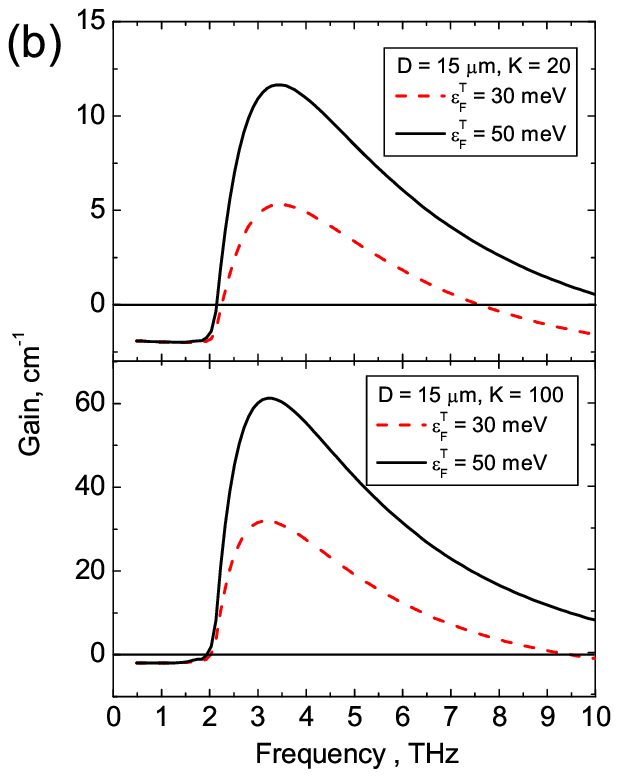}
\caption{Frequency dependences of THz gain in laser structures with different 
width of the slot $D$  and different number of GL $K$ at different values of the quasi-Fermi energy $\varepsilon_F^T$: (a) $D = 10~\mu$m and (b) 
$D = 15~\mu$m.
}
\end{center}
\end{figure}   

\begin{figure}[t]
\vspace*{-0.4cm}
\begin{center}
\includegraphics[width=7.0cm]{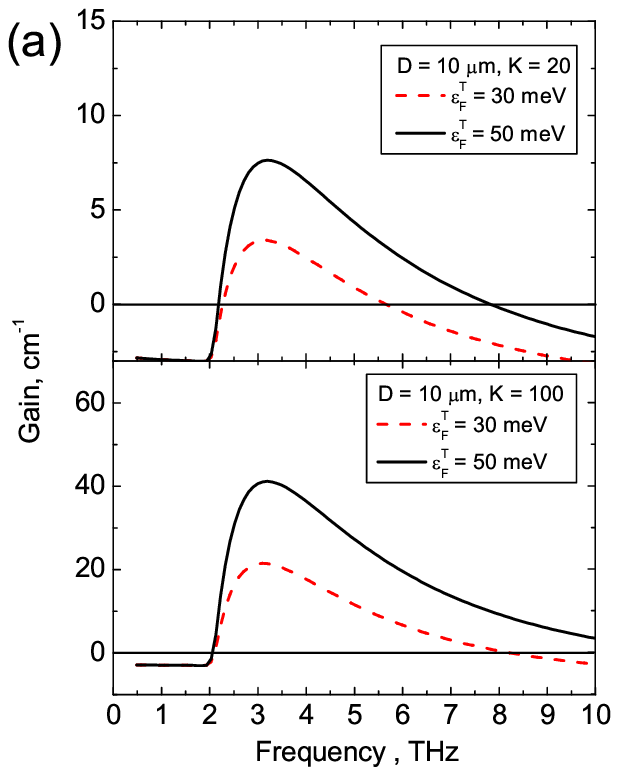}
\includegraphics[width=7.0cm]{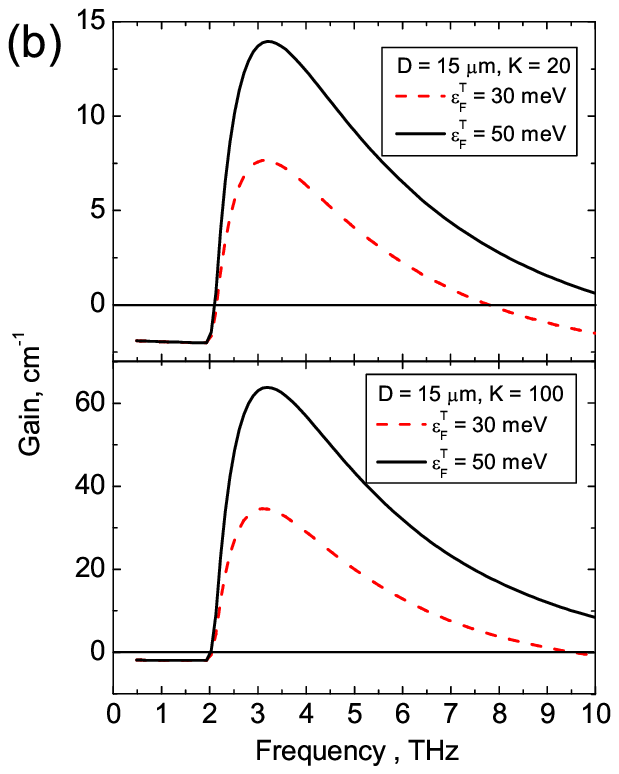}
\caption{The same as in Fig.~4 but for laser  structures without bottom GL.
}
\end{center}
\end{figure}  

Since  the thickness of the MGL structure
is small in comparison with the wavelength of THz  radiation,
the generation and absorption  are determined
by the real part of the net dynamic conductivity
which is the sum of the real parts of the dynamic
conductivity of the bottom GL ${\rm Re}~\sigma_{\omega}^{B}$
and   other GLs $\sigma_{\omega}^{(k)}$: 

\begin{equation}\label{eq2}
{\rm Re}~\sigma_{\omega}
= {\rm Re}~\sigma_{\omega}^{B} + {\rm Re}\sum_{k=1}^K\sigma_{\omega}^{(k)}.
\end{equation}
Considering the expressions for ${\rm Re}~\sigma_{\omega}^{B}$
and ${\rm Re}\sigma_{\omega}^{(k)}$ obtained previously~\cite{11},
one can arrive at the following:
$$
{\rm Re}~\sigma_{\omega} = 
\displaystyle
\biggl(\frac{e^2}{4\hbar}\biggr)\biggl\{
\displaystyle
\frac{4k_BT\tau_B}
{\pi\hbar(1 + \omega^2\tau_B^2)}
\ln \biggl[1 + \exp\biggl(\frac{\varepsilon_F^{B}}{k_BT}\biggr)\biggr]
$$
$$
+ 
\displaystyle
\frac{8k_BT\tau}
{\pi\hbar(1 + \omega^2
\tau^2)}
\sum_{k = 1}^K\ln \biggl[1 + \exp\biggl(\frac{\varepsilon_F^{(k)}}{k_BT}\biggr)\biggr]
$$
\begin{equation}\label{eq3}
+
 \displaystyle
\sum_{k = 1}^K\tanh\biggl(\frac{\hbar\omega - 2\varepsilon_F^{(k)}}{4k_BT}\biggr)
\biggr\}.
\end{equation}
Here  $\tau_B$ and  $\tau$ are the electron and hole momentum relaxation 
times in the bottom and other GLs, respectively,
$T$ is the electron
and hole temperature, and $k_B$ is the Boltzmann constant. 
The first two terms in the right-hand side of Eq.~(3) are associated with 
the intraband (Drude) absorption of THz radiation in all GLs,
whereas the third term corresponds to  the interband transitions.
In the case of lasers without the bottom GL, one can use Eq.~(3) formally setting 
$\tau_B = \infty$. 

The  electromagnetic waves propagating along the SLW was considered
using the following equation which is a consequence
of the Maxwell equations:

\begin{equation}\label{eq4}
\frac{\partial^2E_{\omega}(y,z)}{\partial y^2} + 
\frac{\partial^2E_{\omega}(y,z)}{\partial z^2} +
\biggl[\frac{\omega^2}{c^2}\eta(y,z) - q^2\biggr]E_{\omega}(y,z) = 0.
\end{equation}
Here $E_{\omega}(y,z)$ the amplitude of the 
$y-$th component of the THz electric field $E(t,x,y,z,) = 
E_{\omega}(y,z)\exp[i(qx - \omega\,t)]$, $\eta(y,z)$ is the complex permittivity , $q$ is the wave number of the propagating mode.
The quantities $E_{\omega}(y,z)$, $dE_{\omega}(y,z)/dz$, and
$\eta^{-1}(y,z)d[\eta(y,z)E_{\omega}(y,z)]/dy$ are continuous at the interfaces
between the layers with different refractive indices.
The boundary conditions for the guided mode correspond to the condition 
$E_{\omega}(y,z) \rightarrow 0$ at $y, z \rightarrow \pm\infty$.
Equation~(4) was solved numerically using the effective index and transfer-matrix methods (see, for instance, Refs.~\cite{18,19}).
The coefficient of absorption of the propagating mode $\alpha_{\omega}$
and the coefficient of reflection from the interfaces
between the laser structure edges and vacuum $R$ were calculated using the following formulas, respectively: $\alpha_{\omega} = 2 {\rm Im}~q$ and 
$R = |(q - q_0)/(q + q_0)|^2$, where $q_0 = \omega/c$.
The THz gain, which  describes the attenuation or amplification of the
propagating mode, under optical pumping
can be calculated using the following formula:

\begin{equation}\label{eq5}
g_{\omega }
= \frac{4\pi {\rm Re}\sigma_{\omega}}{c\sqrt{\eta_S}}\Gamma_{\omega}
- \alpha_{\omega},
\end{equation}
where 
\begin{equation}\label{eq6}
\Gamma_{\omega} = \displaystyle
\frac{\int_{-D/2}^{D/2}|E_{\omega}(y,0)|^2dy}
{\int_{-\infty}^{\infty}\int_{-\infty}^{\infty}|E_{\omega}(y,z)|^2dydz}
\end{equation}
is the gain-overlap factor (in the case when $\sigma_{\omega}$ is independent of coordinate $y$) and 
$\eta_S$ is the permittivity of the substate (SiC or Si).

\begin{figure}[t]
\vspace*{-0.4cm}
\begin{center}
\includegraphics[width=7.0cm]{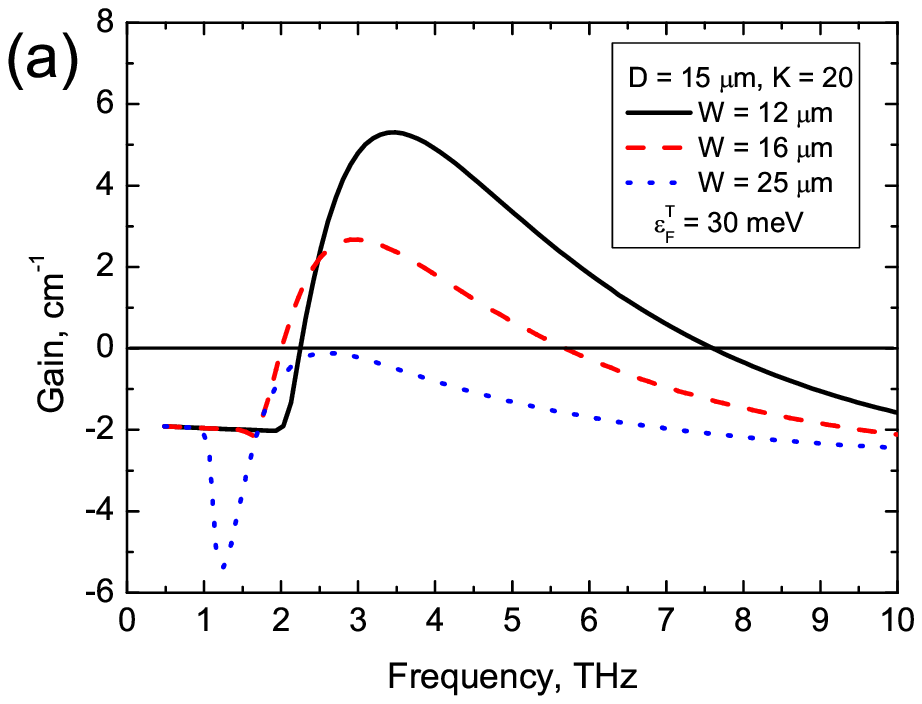}
\includegraphics[width=7.0cm]{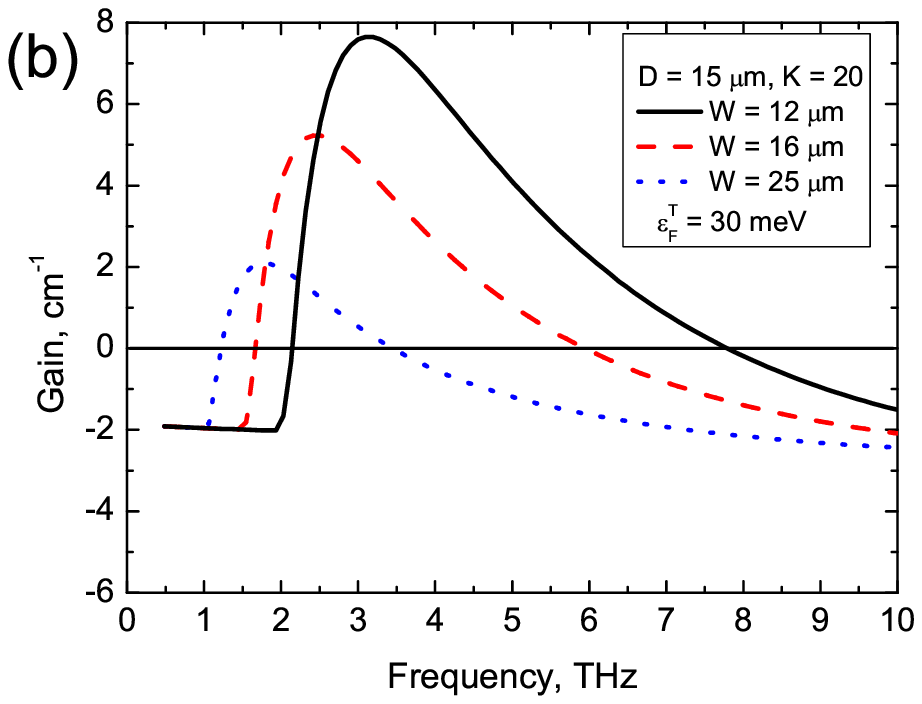}
\caption{Comparison of THz gain versus frequency dependences
in laser structures(a)  with and (b) without bottom GL
and  with different spacing between MGL structure
and back electrode $W$.
}
\end{center}
\end{figure}  
\begin{figure}[t]
\vspace*{-0.4cm}
\begin{center}
\includegraphics[width=7.0cm]{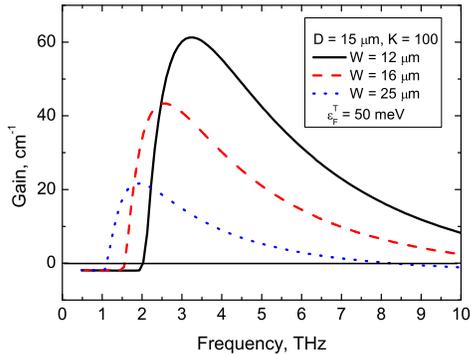}
\caption{The same as in Fig.~6(a) but for $K = 100$
and $\varepsilon_F^T = 50$~meV.
}
\end{center}
\end{figure} 

\section{Results and discussion}

Figure~2 shows the absorption coefficient $\alpha_{omega}$ of
 electromagnetic waves
propagating along the SLW and the coefficient of reflection from 
the laser structure edges
as  function of frequency $\omega/2\pi$
calculated using Eq.~(4) for 
different widths of the slot $D$ and the strips $D_c$ and the substrate
thickness $W$.
The strips and the back electrode are assumed to be made of~Al.
Examples of the spatial distributions of the THz electric field in 
the electromagnetic wave in a laser structures
with $D = 15~\mu$m  at $\omega/2\pi = 1.8$~THz in question are show in Fig.~3.
The electric field distributions 
with the frequency dependences of the dynamic conductivity of
the MGL structures obtained by solving  Eq.~(4)
were substituted to Eqs.~(5) and (6)  to find
the gain-overlap factor and the THz gain.
Figures~4  and 5 show the frequency dependences of the THz gain
$g_{\omega}$
calculated for the laser structures with (Fig.~4) and without
(Fig.~5) bottom GL and  with different structural  parameters
at different pumping conditions (different values of the quasi-Fermi energy)
and at $T = 300$~K. It was assumed that $\hbar\Omega = 920$~meV,
$\tau_B = 1$~ps, and $\tau = 10$~ps.
As seen from comparison of Figs.~4 and 5, the laser structures
with and without the bottom GL exhibit  qualitatively
similar frequency dependences
of the THz gain $g_{\omega}$ (at chosen geometrical parameters) but with
somewhat higher maxima in the latter structures.
The frequency $\omega_{min}$ at which
$g_{\omega}$ changes its sign is  $\omega_{min}/2\pi \simeq 2$~THz
in the laser structures of both types.
The latter
 is different from the situation in the MGL lasers with the Fabri-Perot
resonator~\cite{11} in which eliminating the bottom GL can lead to
a pronounced decrease in $\omega_{min}$. This is attributed to
the effect of SLW: the variation of the frequency results in a redistribution
of the spatial distribution of the THz electric field and, hence, in a change
in the gain-overlap factor. 
Due to this,
the spacing between the MGL structure $W$ and the back electrode affects the
the frequency dependences of the THz gain.
Figure~6 shows these dependences calculated for different $W$.
One can see that the height of the THz gain maxima decreases with increasing 
$W$. Simultaneously the frequency $\omega_{min}$  shifts toward lower values. 
This shift is more pronounced in the laser structures without the bottom GL
[compare Figs.~6(a) and 6(b)].
As seen from Fig.~6(b), $\omega_{min}/2\pi$
can reach 1~THz. In the case of larger $K$ and $\varepsilon_F^T$,
as seen from Fig.~7,  the THz gain in the range 
$\omega/2\pi \gtrsim 1 - 2$~THz can be rather large.

A decrease in the momentum 
relaxation time $\tau$ (which results in an enhancement of the 
Drude absorption) leads to an increase in $\omega_{min}$ and 
to a decrease of $|{\rm Re}\sigma_{\omega}|$ in the range of frequencies where
Re~$\sigma_{\omega} < 0$. 
This is seen in 
 Fig.~8.

The above results correspond to $T = 300$~K. 
Lowering of the temperature should lead to widening of the frequency
range where Re~$\sigma_{\omega}$ is negative
 and where $g_{\omega}$ can be positive.
As a result, $\omega_{min}$ might decrease with decreasing temperature.
The latter is confirmed by the calculated temperature dependences of 
$\omega_{min}$ shown in Fig.~9.

\begin{figure}[t]
\vspace*{-0.4cm}
\begin{center}
\includegraphics[width=7.5cm]{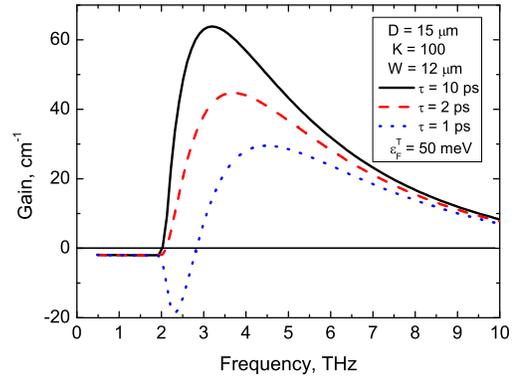}
\caption{Frequency dependences of THz gain in laser structures (without
the bottom GL) with different $\tau$.
}
\end{center}
\end{figure} 

\begin{figure}[t]
\vspace*{-0.4cm}
\begin{center}
\includegraphics[width=7.0cm]{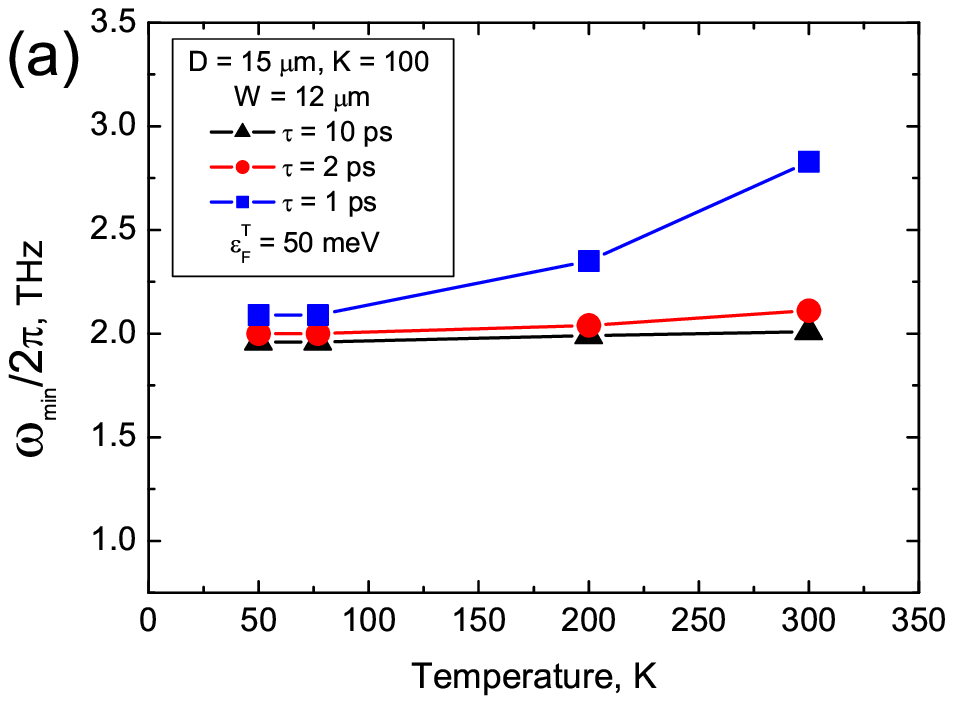}
\includegraphics[width=7.0cm]{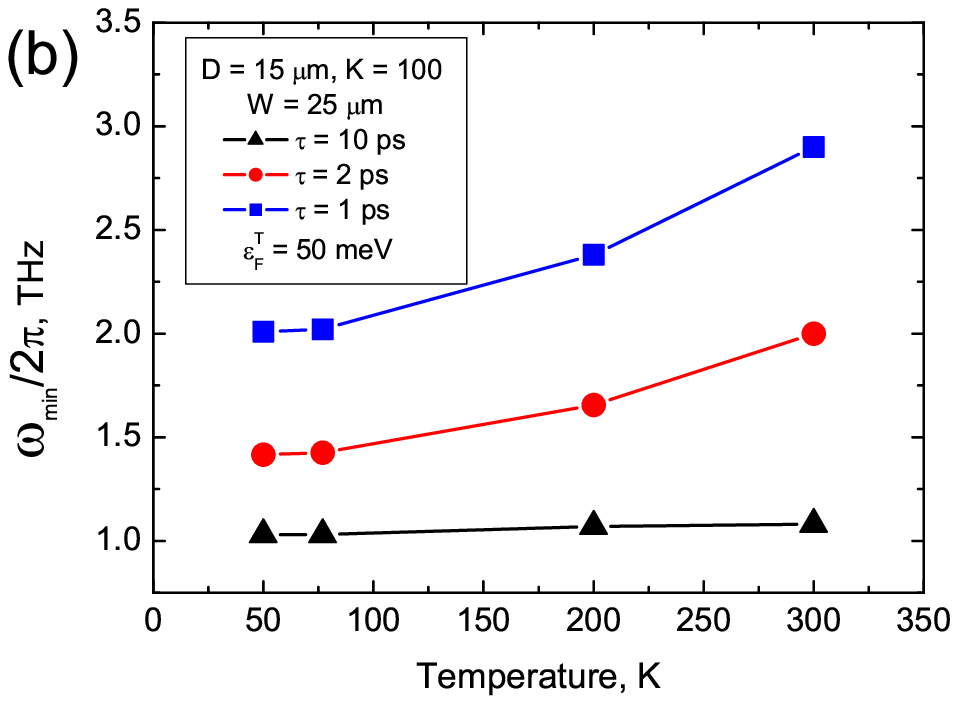}
\caption{Temperature dependences of $\omega_{min}$ for laser structures with different $\tau$: (a) $W = 12~\mu$m and (b) $W = 25~\mu$m.
}
\end{center}
\end{figure}  

 \begin{figure}[t]
\vspace*{-0.4cm}
\begin{center}
\includegraphics[width=7.0cm]{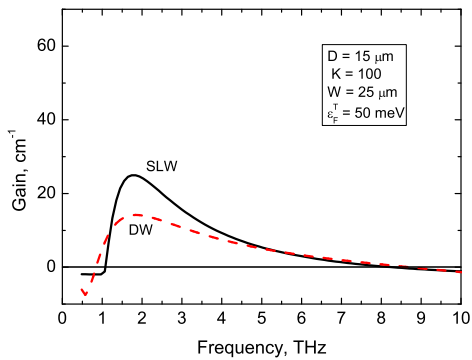}
\caption{Comparison of the frequency dependences
of THz gain in laser structures with
SLW and DW
}
\end{center}
\end{figure}

Figure~10 shows the frequency dependences of the THz gain calculated for
laser structures with SLW and DW for $T = 300$~K and $\tau = 10$~ps.
One can see that the maximum of the THz gain in the laser with SLW is
somewhat higher than that in  the laser with DW. This can primarily
be explained
by the effect of the spatial distribution of the THz electric field on
the gain-overlap factor [compare Figs.~3(a) and 3(b)]. To maximize the THz gain at 
the desirable frequency one needs carefully optimize the geometrical parameters.

Considering the propagation of the THz electromagnetic wave in the laser
structure (with the length $L$ in the $x$-direction)
  and its reflection from edges or from the external mirrors
(with the reflective coefficients $R_1$ and $R_2$), 
the condition of lasing can be presented as

\begin{equation}\label{eq7}
L\,g_{\omega} > \ln\frac{1}{R_1R_2}.
\end{equation}
For a laser structure with SLW with $D = 15~\mu$m, $W = 12~\mu$m,
and $K = 100$ setting $\varepsilon_F^T = 30 - 50$~meV
and $\omega/2\pi = 3.0$~THz, so that $R_1 = R_2 \simeq 0.05$
and $g_{\omega} \simeq 30 - 60$~cm$^{-1}$ (see Figs.~2(b) and 4(b),
respectively), one obtains
$L > 0.1 - 0.2$~cm.
If $D = 15~\mu$m, $W = 25~\mu$m,
and $K = 100$ at $\varepsilon_F^T = 50$~meV
and $\omega/2\pi = 1.5$~THz ($R_1 = R_2 \simeq 0.05$
and $g_{\omega} \simeq 10$~cm$^{-1}$ as follows from
 Figs.~2(b) and 7,
respectively), one obtains
$L > 0.6$~cm.

As shown above, at the quasi-Fermi in the topmost GL
about $\varepsilon_F^T =30 - 50$~meV, the achievement of the THz lasing in the devices under consideration
 at room temperatures is feasible.
At $T = 300$~K
the condition $\varepsilon_F^T \gtrsim 30$~meV corresponds to
the electron and hole densities about $2\times10^{11}$~cm$^{-2}$.
Such densities can be obtained at reasonable optical powers(see Ref.~\cite{12}
and the references therein).

\section{Conclusions}
We proposed THz lasers with optical pumping based on MGL structures with
waveguides and calculated their characteristics. The feasibility of  lasing in the devices under consideration
at the low end of  the THz frequency range
at room temperatures  was demonstrated. 

\section*{Acknowledgments}
The authors are grateful to V.~V.~Popov,
M.~Ryzhii, A.~Satou, M.~Suemitsu, and F.~T.~Vasko
for fruitful discussions.
This work was supported by the Japan Science and Technology Agency, CREST, 
Japan.


\begin{thebibliography}{99}
\bibitem{1}
A. H. Castro Neto, F. Guinea, N. M. R. Peres, 
K. S. Novoselov, and A. K. Geim,
Rev. Mod. Phys. {\bf 81},  109 (2009).


\bibitem{2}
F.~T.~Vasko and V.~Ryzhii,
Phys. Rev B {\bf 77}, 195433 (2008).
%
\bibitem{3}
V.~Ryzhii, V.~Mitin, M.~Ryzhii, N.~Ryabova, and T.~Otsuji,
Appl. Phys. Exp. {\bf 1}, 063002 (2008).
\bibitem{4}
F.~Xia, T.~Mueller, R.~Golizadeh-Mojarad, M.~Freitag, Y.~Lin,
J.~Tsang, V.~Perebeinos, and P.~Avouris,
Nano Lett. {\bf 9}, 1039  (2009).

\bibitem{5}
V. Ryzhii and M. Ryzhii, 
Phys. Rev. B {\bf 79},    245311  (2009). 
\bibitem{6}
Y.~Kawano and K.~Ishibashi,
Proc. 34th Int. Conf. on Infrared, and Terahertz Waves,
Busan, Korea, Sept.21-25, 2009, W4A04.0380.

\bibitem{7}
V.~Ryzhii, M.~Ryzhii, and T.~Otsuji,
J. Appl. Phys.  {\bf 101},  083114 (2007).
\bibitem{8}
F.~Rana,
IEEE Trans. Nanotechnol. {\bf 7}, 91 (2008).
\bibitem{9}
A.~Satou, F.~T.~Vasko, and V.~Ryzhii, 
Phys. Rev. B {\bf 78},  115431  (2008). 




\bibitem{10}
A.~A.~Dubinov, V.~Ya. Aleshkin, M.~Ryzhii, T.~Otsuji, and V.~Ryzhii,
Appl. Phys. Exp. {\bf 2},  (2009).

\bibitem{11}
V.Ryzhii, M.~Ryzhii, A.~Satou, T.~Otsuji, A.~A.~Dubinov, and V.~Ya.~Aleshkin,
J. Appl. Phys., {\bf 106}, 084507 (2009).

\bibitem{12}
V.~V.~Cheianov and V.~I.~Fal'ko,
Phys. Rev. B {\bf 74}, 041403(R) (2007).

\bibitem{13}
M.~Ryzhii and V.~Ryzhii,
Jpn. J. Appl. Phys.  {\bf 46}, L151 (2007).

\bibitem{14}
M.~Ryzhii and V.~Ryzhii,
Physica E, {\bf 40}, 317 (2007).



\bibitem{15}
P.~Neugebauer, M.~Orlita, C.~Faugeras,  A.-L.~Barra, and M.~Potemski,
Phys. Rev. Lett. {\bf 103}, 136403 (2009).
\bibitem{16}
A.~Bostwick, T.~Ohta, T.~Seyller, K.~Horn, and E.~Rotenberg,
Nature Phys. {\bf 3}, 36 (2007). 

\bibitem{17}
F.~Varchon, R.~Feng, J.~Hass, X.~Li, B.~Ngoc Nguyen, C.~Naud, P.~Mallet, J.-Y.~Veuillen, C.~Berger, E.~H.~Conrad, and L.~Magaud,
Phys. Rev. Lett. {\bf 99}, 126805 (2007).

\bibitem{18}
K.~J.~Ebeling, 
{Integrated Optoelectronics: Waveguide Optics, Photonics, Semiconductors} 
(Springer-Verlag, Berlin, 1993).

\bibitem{19}
M.~Born and E. Wolf, { Principles of Optics: Electromagnetic Theory of 
Propagation, Interference and Difraction of Light} (Pergamon Press, Oxford, 1964).


%

\end{thebibliography}
\end{document}